\documentclass[12pt,preprint]{aastex}

\def\m{$\mu$m}
\def\ab{$\sim$}

\def\lya{Ly$\alpha$}
\def\Lya{Ly$\alpha$ }

\def\ab{$\sim$}
\def\deg{\ifmmode ^{\circ}
         \else $^{\circ}$\fi}

\shorttitle{}
\shortauthors{Colbert et al.}

\begin{document}


\title{Ultraviolet-Bright, High-Redshift ULIRGS}


\author{James W. Colbert\altaffilmark{1}, Harry Teplitz\altaffilmark{1},
Paul Francis\altaffilmark{2}, Povilas Palunas\altaffilmark{3}, Gerard 
M. Williger\altaffilmark{4,5}, Bruce Woodgate\altaffilmark{6}}

\altaffiltext{1}{Spitzer Science Center, California Institute of Technology,
    Pasadena, CA 91125}

\altaffiltext{2}{Research School of Astronomy and Astrophysics, The Australian National University, 
    Canberra, ACT 0200, Australia}

\altaffiltext{3}{McDonald Observatory, University of Texas, Austin, TX 78712}

\altaffiltext{4}{Dept. of Physics \& Astronomy, University of Louisville, Louisville, KY 40292}

\altaffiltext{5}{Dept. of Physics \& Astronomy, John Hopkins University, Baltimore, MD 21218}

\altaffiltext{6}{NASA Goddard Space Flight Center, Greenbelt, MD 20771}

\begin{abstract}

We present {\it Spitzer} Space Telescope observations of the $z$=2.38 \lya-emitter over-density 
associated with galaxy cluster J2143-4423, the largest known structure (110 Mpc) 
above $z=2$. 
We imaged 22 of the 37 known \lya -emitters within the filament-like structure, 
using the MIPS 24\m\ band. We detected 6 of the  
\lya -emitters, including 3 of the 4 clouds of extended ($>$50kpc) \Lya emission, also known
as \Lya Blobs. 
Conversion from rest-wavelength 7\m\ to total far-infrared luminosity 
using locally derived correlations suggests all the detected sources are in the class of
ULIRGs, with some reaching Hyper-LIRG energies.  
\Lya blobs frequently show evidence for interaction, either in {\it HST} imaging, 
or the proximity of multiple MIPS sources within the \Lya cloud. This connection suggests
that interaction or even mergers may be related to the production of \Lya blobs. A connection
to mergers 
does not in itself help explain the origin of the \Lya blobs, as most of the
suggested mechanisms for creating \Lya blobs (starbursts, AGN, cooling flows) could also 
be associated with galaxy interactions.

\end{abstract}


\section{Introduction}
  
The 110 Mpc filament mapped out by 37 Ly$\alpha$-emitting objects around the $z$ = 2.38 galaxy 
cluster J2143-4423 is the largest known structure above $z=2$ \citep{pal04},
comparable in size to some of the largest structures seen in the local Universe 
\citep[i.e. the Great Wall,][]{gel89}. Initially identified from 
narrow-band imaging tuned to \Lya \ at $z$=2.38, it has since been spectroscopically confirmed 
\citep{fra04,frb04}. 
In addition to its compact \lya -emitters, this high-redshift ``Filament'' is also home to 
four extended \lya -emitting clouds, known more commonly as Lyman $\alpha$ blobs. 

The Ly$\alpha$ blob is a relatively new class of objects found among high-redshift galaxy 
over-densities \citep{ste00,kee99,pal04}. While similar in extent ($\sim$100 kpc) and \Lya flux 
(\ab 10$^{44}$ ergs s$^{-1}$) to high-redshift radio galaxies, 
blobs are radio quiet and are therefore unlikely to arise from interaction with jets. 
Current surveys have reported the discovery of roughly 10 of these giant \Lya blobs, but they 
are not isolated high redshift oddities. \cite{mat04} have demonstrated that the blobs are 
part of a continuous size distribution of resolved ($>$16 arcsec$^2$) \Lya emitters, with more 
than 40 presently known.

One of the standing mysteries of the blobs is the source of their energy, as the measured 
ultraviolet flux from nearby galaxies is insufficient to produce the observed
\Lya fluxes. 
One possibility is that the \Lya blobs are powered by supernova-driven superwinds 
\citep{ohy04}, driving great plumes of gas into the surrounding ambient medium and producing 
shocks. An obscured AGN is another model, with the exciting ultraviolet illumination escaping along different
lines of sight \citep[i.e.,][]{bas04}. Cooling flows  
have also been suggested \citep{fard01,fra01} 
as a possible power source. 

There is growing evidence that \Lya blobs mark regions of extreme infrared 
luminosity. Submillimeter flux has been detected in two of the giant \Lya blobs,  
SMM J221726+0013 \citep{chp01} and SMM J17142+5016 \citep{sma03}. 
The submm source SMM J02399-0136 is also likely surrounded by a \Lya blob halo, as its 
\Lya emission covered most of a 15$\arcsec$ slit \citep{ivi98}. Most recently, 
\cite{gea05} has detected submm flux from four of the smaller ($<$ 55 square arcsec) 
and less luminous ($<$ 2$\times$10$^{43}$ L$_{\odot }$) \Lya blobs from \cite{mat04}: 
LAB5, LAB10, LAB14 \& LAB18. 
Also, \cite{dey05} has discovered a single \Lya blob (SST24 J1434110+331733) in the NOAO Deep Wide-Field 
Survey with strong 24\m\ flux (0.86 mJy).

In this paper we discuss the {\it Spitzer} 24\m\ observations of these $z$=2.38 \lya -emitters, both
compact sources and blobs. 
We estimate the total far-infrared luminosity for all detections and discuss
the possibility of a connection between mergers and \Lya blobs. 
We assume an $\Omega _{M}$=0.3, 
$\Omega _{\Lambda }$=0.7 universe with H$_o$=70 km s$^{-1}$ Mpc$^{-1}$.

\section{Observations}

Our data were obtained using Multiband Imaging Photometer for Spitzer \citep[MIPS;][]{rie04} 
in 24\m\ photometry mode. The field was observed on October 13 and 14 and November 4 and 5, 2004.
The primary observation was a 3$\times$5 raster map, covering approximately 15$\times$25
arcminutes of the filament structure, centered at 21$^h$42$^m$38.0$^s$, -44$\deg $26$\arcmin$30$\arcsec$. 
The total integration time per pixel is 1818 seconds. 
In addition we observed one small side field (5$\times$5 arcmin) with a 
single known \lya -emitter (at 21$^h$42$^m$56.3$^s$, -44$\deg $37$\arcmin$57$\arcsec$) for 666 seconds. 
The data were initially reduced by the standard {\it Spitzer} 
pipeline \footnotemark .
We then produced sky frames for each data frame from the median of the 100 nearest 
(in time) data frames.

\footnotetext{http://ssc.spitzer.caltech.edu}

We assembled the final mosaic of the MIPS 24\m\ image using the MOPEX package available from the
{\it Spitzer} Science Center.
The final drizzled pixel scale is 1.23$\arcsec$ per pixel. For source extraction, we applied the
APEX source extraction package available within MOPEX. APEX requires an input Point Response Function 
(PRF), which we created from the 40 brightest objects extracted from the final image.
Our final APEX source extraction detected 3010 objects down to
roughly 50 $\mu$Jy. The full dataset will be discussed in a future paper.

The 5-$\sigma$ detection limit, as measured within a 7.5$\arcsec$ aperture, is 58 $\mu$Jy
over most of the image, but can be as high as 180 $\mu$Jy at the edges. 
None of the \lya -emitters for which APEX failed to find MIPS sources had aperture fluxes greater 
than these 3-$\sigma$ upper limits. 
A list of all extracted MIPS sources associated with \lya -emitters is presented in Table 1. 
A brief note on the naming conventions: The first three \lya -emitters were found by \cite{fra96}
and named B1, B2, \& B3. Further study by \cite{fra97} demonstrated that the B3 source was not real, 
but they also found another source, which they labeled B4. 
\cite{pal04} found 34 additional \lya -emitters in the field and so, rather than label each one, 
only gave B\# designations to the three new \Lya blobs: B5, B6, \& B7. 
We continue the B\# naming convention, adding sources B8 and B9.

\section{MIPS Detections of \Lya Sources}

Six out of 22 ($\sim$30\%) of the Ly$\alpha$-emitters within the filament are 
associated with MIPS 24\m\ sources (Figure 1), including three of the four resolved 
($>$50 kpc) Ly$\alpha$ blobs. 
This blob association rate is much higher than that for unresolved 
Ly$\alpha$-emitters (Only 3 of 18). We determined a MIPS source to be associated if it
lies within one MIPS FWHM (5.9$\arcsec$) of the position of the Ly$\alpha$-emitter.
In four cases the coordinates for the MIPS and \Lya \ sources agree to less 
than an arcsecond, while the remaining two are separated by 2.7 (Blob B7) and 4.1$\arcsec$ 
(source B9).
We checked the MIPS field astrometry against that of the optical data by comparing object centroids
for all detected MIPS sources against those found in I-band. After correction for a 0.3$\arcsec$
offset, we found that the positions of the objects agree with a standard deviation of
1.5$\arcsec$. 

The density of MIPS sources in the central portion of the field is \ab 4.5 arcmin$^{-2}$, making 
the odds of a chance superposition of a random MIPS source $<$0.4\% for those within an arcsecond, 
$<$ 2.9\% for within 2.7$\arcsec$, and $<$ 6.6\% for within 4.1$\arcsec$. However, all the
detected Ly$\alpha$-emitter counterparts have fluxes greater than 120 $\mu$Jy, well above the
detection limits and with a consequently lower surface density. Accounting for source brightness 
decreases the probability of chance superposition below 0.3\% for all sources except for 
source B9, which has a 3.3\% chance that its associated MIPS source 
(4.1$\arcsec$ away) is mere coincidence.
For comparison, the \cite{dey05} MIPS source  lies 2.5$\arcsec$ 
from the center of its associated \Lya blob.

We do not detect 16 of the 22 Ly$\alpha$-emitters in the MIPS image. Six of these sources lie in
regions of higher noise, where the 5$\sigma$ detection limits range from 100-180 $\mu$Jy.
The remaining ten Ly$\alpha$-emitters are in the central regions of the image, with 
detection limits of $<$60 $\mu$Jy. The lack of faint MIPS counterparts is unlikely to be a 
result of error in object positions, as we examine a radius nearly 4 times the 
typical 1$\sigma$ offset found from matching the MIPS and optical data.
The absence of detections between 60 and 120 $\mu$Jy indicates there 
is not a rapid climb in numbers with decreasing 24\m\ flux for Ly$\alpha$-emitters over this range, 
although the number of objects considered is small. 

Two of the Ly$\alpha$ blobs, B6 and B7, appear to have additional MIPS sources associated with 
them. The B6 Ly$\alpha$ blob extends over 25$\arcsec$, or over 200 kpc at $z$=2.38, and
contains more than one knot or concentration. The brightest Ly$\alpha$ knot is the central one
and it is associated with a bright MIPS source, but the southern knot (not to be confused with 
the large diffuse area to the southeast) also has a MIPS source of almost identical brightness. 
There is also an area of more diffuse Ly$\alpha$ emission to the north which also appears to be 
associated with a bright MIPS source. Each source is separted by approximately 9.5$\arcsec$ from 
the central source, creating a potential triple system. The northern source should probably be 
treated with some caution, as it not only lacks any clear Ly$\alpha$ concentration, but there
is also a relatively bright galaxy visible at that location in the B-band image (B=22.5), which is 
unlikely to be at high redshift. 
 
The center of the B7 Ly$\alpha$ blob actually lies between two sources. We have associated 
it with the brighter, closer MIPS source 2.7$\arcsec$ (\ab 20 kpc) to its southeast, but a second 
source lies 5.9$\arcsec$ (\ab 45 kpc) to the northwest. While there is a \ab 10\% chance the second 
object could be chance association, the location of the Ly$\alpha$ blob immediately between the 
two MIPS sources suggests a possible physical connection. 

\section{Ultraviolet-Bright ULIRGs}

The detected 24\m \ flux densities range from 0.1 to 0.6 mJy, which at $z$=2.38 corresponds to 
roughly 2$\times 10^{11}$ to 10$^{12}$ L$_{\sun }$ in rest frame 7\m\ $\nu F\nu$. To convert from the
mid-infrared to total bolometric luminosity (L$_{bol}$), we use the relationship from 
\cite{cha01}, hereafter CE01, calibrated by examining galaxies measured with both ISO and IRAS:

\begin{equation}
L_{IR} = 4.37^{+2.35}_{-2.13} \times 10^{-6} \times L^{1.62}_{6.7\mu m} 
\end{equation} 

In an effort to avoid over-estimation of L$_{IR}$, the mid-infrared conversion 
used throughout this paper assumes the low end of the 1$\sigma$ envelope, producing total 
luminosities roughly a factor of 2 lower than a direct application of the formula.
Even this conservative mid-IR conversion puts all the sources
in the class of ultra luminous infrared galaxies (ULIRGs; $> 10^{12} L_{\sun }$), with many
achieving Hyper-LIRG ($> 10^{13} L_{\sun }$) status (see Figure 2).  
It should be noted, however, that the CE01 relation is based on nearby 
starburst-dominated galaxies and it is possible it may not hold at high redshift or for the 
most extreme starbursts, if their spectral energy distributions are significantly different.

The L$_{bol}$ of most ULIRGs in the nearby universe comes from a combination of strong star 
formation and AGN activity \citep[see][for review]{san96}. 
The majority are dominated by the stellar component, seen both in the far-infrared to radio 
relation \citep{yun01} and their strong PAH to continuum ratios \citep{lut98}. However, 
this changes as the luminosity approaches that of the Hyper-LIRG (LIR$>$10$^{13}$ L$_{\odot }$); AGNs 
typically dominate above \ab 10$^{12.5}$ L$_{\odot }$ \citep{tra01}. 
It is presently unknown how this trend will continue out to higher redshifts, where high-mass
mergers, like those predicted to build giant ellipticals, could be producing energies on the scale 
of a Hyper-LIRG through star formation \citep{dop05}. PAHs are already being found in high-redshift
infrared sources \citep{hou05,yan05}.  
If these MIPS sources are AGN-dominated, one would expect $F_{\nu } \propto \nu ^{-1}$, which would
still place all the detected sources in the class of ULIRGs. 

If we combine the 24\m\ detections from this study and from \cite{dey05} with
the submm detections of \Lya blobs \citep{chp01,sma03,gea05}, 
there are now at least 10 known high redshift ULIRGs surrounded by \Lya halos.
We plot \Lya luminosity versus L$_{bol}$ for all the identified \Lya 
blobs with infrared or submm detections in Figure 3. SMM J02399-0136 \citep{ivi98}
is excluded as its total \Lya flux is not precisely known.
We use the published L$_{bol}$ values for all submm sources, but derive
the total L$_{bol}$ for the \cite{dey05} 24\m\ source using 
the CE01 relation to better compare with our results. 

Applying a similar analysis, \cite{gea05} found a weak \lya /IR relation for 
submm-detected \Lya blobs with a typical L$_{Ly\alpha }$/L$_{bol}$ ratio just 
under 0.1\%. Including our 24\m -detected \Lya blobs, which are brighter
in \Lya than most of the \cite{gea05} objects, we continue to see the same weak
trend, with a L$_{Ly\alpha }$/L$_{bol}$ efficiency of 0.05-0.2\%. Neither the detection of 
X-rays (SMM J17142+5016 \& LAB18) nor CIV emission (B1, SMM J17142+5016, \&  
SST24 J1434110+331733), both strong indicators of AGN activity, has any clear effect 
on this trend. 
There is one significant outlier, the B5 blob, which falls at least a factor of 30 away
from this proposed relation. While this relation is still tentative, it suggests
a direct causal connection between the ULIRG infrared luminosity and
the \Lya blobs. 

\begin{deluxetable}{ccccccl}
\setlength{\tabcolsep}{0.05in} 
\tablehead{
\colhead{RA} &
\colhead{Dec} &
\colhead{log L$_{Ly\alpha }$} &
\colhead{24\m\ F$_{\nu}$} &
\colhead{$\nu L_{\nu }$} &
\colhead{Total FIR} &
\colhead{Comment} \\
& & ergs s$^{-1}$ & $\mu$Jy & 10$^{10}$ L$_{\odot }$ & 10$^{10}$ L$_{\odot }$ &
}
\tablecaption{\Lya Sources}
\startdata
21:42:27.59  & -44:20:29.7 & 43.9 & 236 & 33.4 & 1000 & Blob B1 \\
21:42:32.16  & -44:20:18.5 & 43.0 & 562 & 79.5 & 4300 & B4 \\
21:42:42.66  & -44:30:09.4 & 43.8 & 634 & 89.5 & 5200 & Blob B6 \\
21:42:34.95  & -44:27:08.8 & 43.5 & 292 & 41.3 & 1500 & Blob B7; 2.7$\arcsec$ offset \\
21:43:05.84  & -44:27:20.9 & 43.7 & 142 & 20.1 & 460 & B8 \\
21:43:37.33  & -44:23:56.2 & 43.1 & 123 & 17.4 & 360 & B9; 4.1$\arcsec$ offset \\
21:43:03.57  & -44:23:44.2 & 43.8 & $<$ 58 & $<$ 8 &  $<$ 80 & Blob B5 \\
\hline
\multicolumn{7}{c}{Possible \Lya Blob Companions} \\
\hline
21:42:42.79  & -44:30:18.2 & n/a & 535 & 76 & 3900 & 9.5$\arcsec$ from B6 \\
21:42:42.49  & -44:29:59.8 & n/a & 340 & 48 & 1900 & 9.5$\arcsec$ from B6 \\
21:42:34.68  & -44:27:01.2 &  n/a & 172 & 24 & 620 & 5.9$\arcsec$ from B7
\enddata
\end{deluxetable}

\section{\Lya Blob Merger Connection?}

The high rate of MIPS detection for the \Lya blobs (3 out of 4) demonstrates that these ultraviolet sources 
are locations of tremendous infrared energy. Two of the \Lya blobs (B6 \& B7) are associated
with multiple MIPS sources, suggesting possible interaction or even merger. These separations between 
possible ULIRG component galaxies are large (60-70 kpc) compared to the typical 
low-redshift ULIRG, which have a median separation \ab 2 kpc \citep{mur96}. 
However, interacting ULIRGs with wide separations on the scale of 50 kpc
do exist \citep{kim02}, and seem to occur when at least one the interacting galaxies is highly 
gas-rich \citep{din01}. 
  
Previous high resolution {\it HST} imaging also indicates a possible \Lya blob connection to 
merging galaxies. For instance, NICMOS imaging of blob B1 shows a pair of compact, red galaxies 
with a projected separation of less than 7 kpc \citep{fra01}. 
Only three other \Lya blobs have been imaged with {\it HST}: the blobs in the
field of 53W002, one of which shows a second nearby galaxy \citep{kee99}, and
blob SSA22a-C11 \citep{ste00} which STIS imaging shows to be composed of several likely 
interacting components \citep{cha04}. \cite{dey05} also find multiple components within their 
blob, visible in their ground-based, optical imaging.  
 
In the local universe, the most energetic mergers are associated with the ULIRGs, with
roughly 90\% of ULIRGs clearly interacting \citep{bus02,far01}. 
The majority of local ULIRG mergers are between two low mass (0.3-0.5 L$^*$) galaxies 
\citep{col01}, but major mergers of high-mass galaxies, like those predicted to build giant 
ellipticals, must be occurring at higher redshift \citep[i.e.,][]{con05}. 
If merger-induced star formation is the source of the majority of the infrared flux in these objects, it
implies star formation rates of 1000s of M$_{\odot }$ yr$^{-1}$, like those suggested
for some sub-mm sources \citep{cha03,gea05}.
Such massive SFRs would be capable of generating the supernova kinetic energy needed to drive a superwind 
and power the \Lya blobs. If supernovas deposit roughly 10$^{49}$ ergs per solar mass of stars 
into the surrounding medium \citep[i.e.,][]{bow01}, a star formation rate of 1000 M$_{\odot }$ yr$^{-1}$
will easily match the \ab 10$^{51}$ ergs yr$^{_1}$ emitted by the blobs, even at a low efficency.

Mergers could also help drive gas inwards into a supermassive black hole, making AGN another
viable infrared energy source, with escaping ultraviolet radiation driving the \Lya blobs.
\cite{dey05} found that their 24\m -detected blob is better fit by an AGN spectral energy distribution 
than that of a star forming galaxy, but longer rest-wavelengths or a mid-IR spectrum are 
required for a definitive determination. There is one clearly starbursting galaxy at the 
northern end of their \Lya nebula. 

Cooling flows also remain a possibility, as it might be common to find multiple galaxies merging
at the center of such a large inward flow of gas. It does become a less favored model, however, as the energy
from the cooling gas in no longer required to power the \Lya blobs. 
The ULIRGs would appear to have more than enough energy in their budget to do so on their own.

This work is based on observations made with the {\it Spitzer} 
Space Telescope, which is operated by the 
Jet Propulsion Laboratory, California Institute of Technology, under NASA contract 1407.
We wish to acknowledge financial support from the {\it Spitzer} grant GO-3699.

\begin{figure}
\epsscale{1.0}
\plottwo{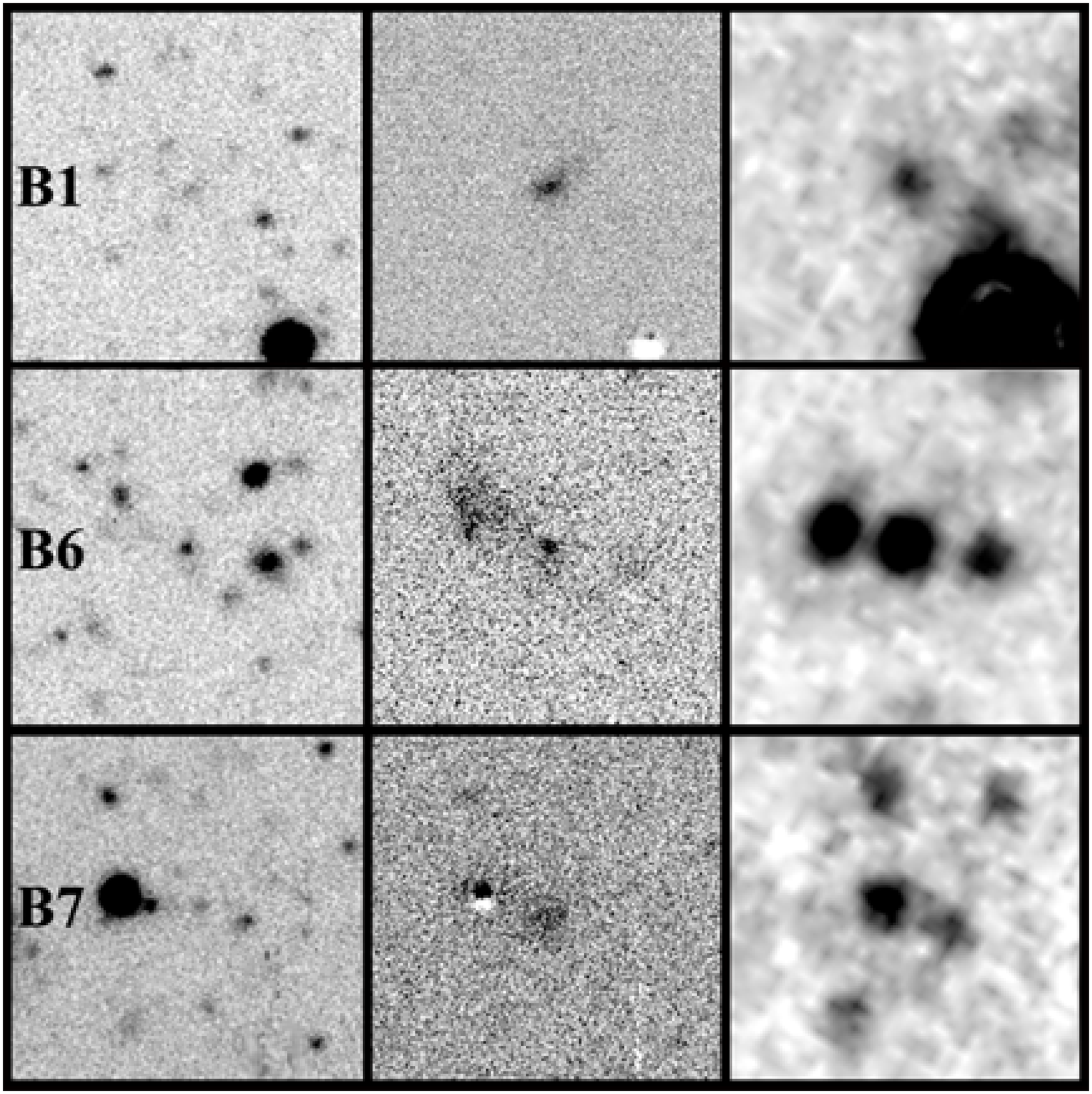}{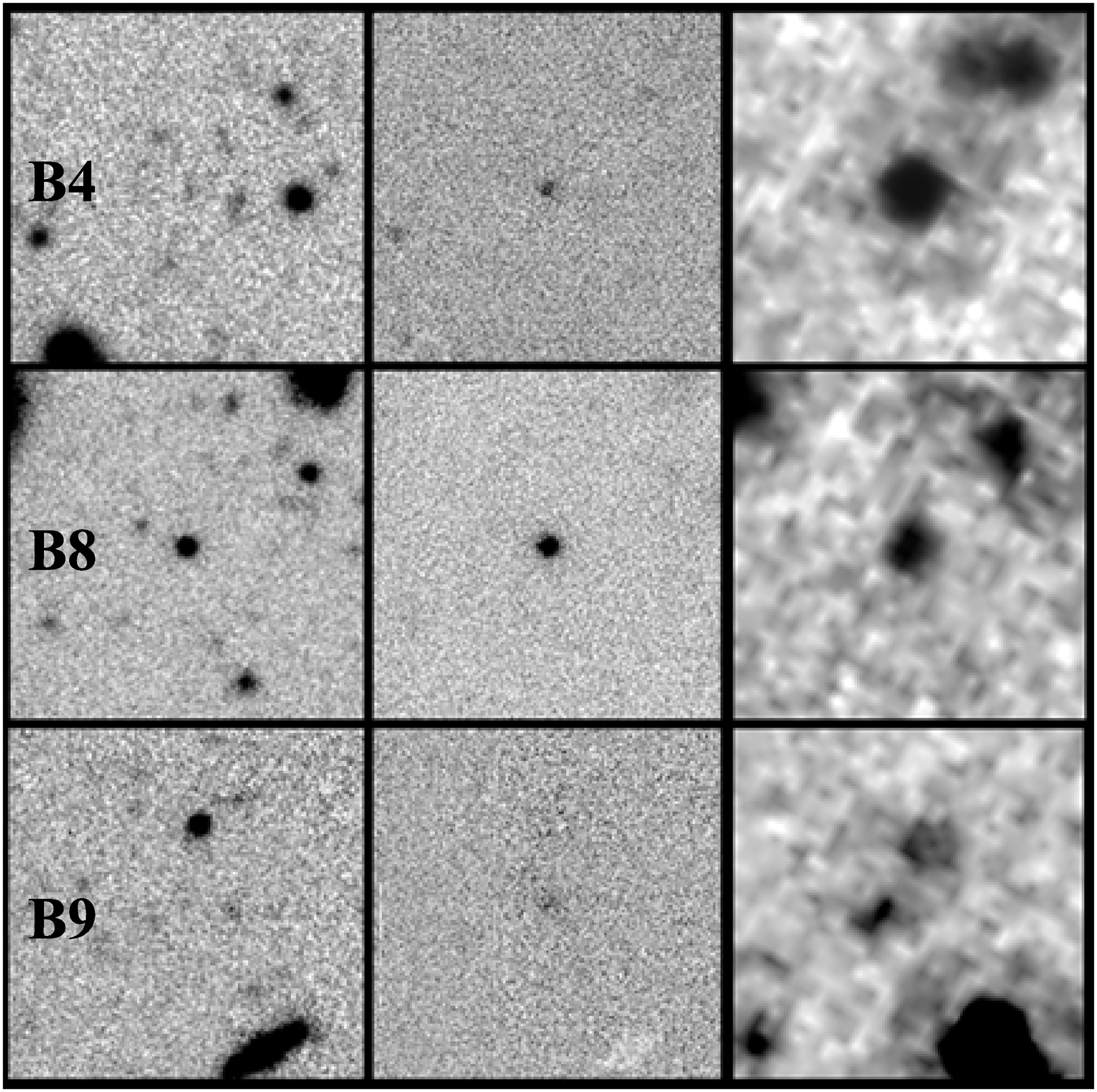}
\caption{Left) Images of the 40$\arcsec$$\times$40$\arcsec$ field around the three MIPS-detected \Lya  
blobs: B1, B6, \& B7. From left to right the images are CTIO B-band, CTIO continuum-subtracted \lya , 
and MIPS 24$\mu$m. East is upwards, north to the right. One arcsecond corresponds to 8kpc 
at z=2.38. Right) Images of the 40$\arcsec$$\times$40$\arcsec$ field around the three MIPS-detected 
non-extended \lya -emitters. Format is same as left.} 
\end{figure}

\begin{figure} 
\plotone{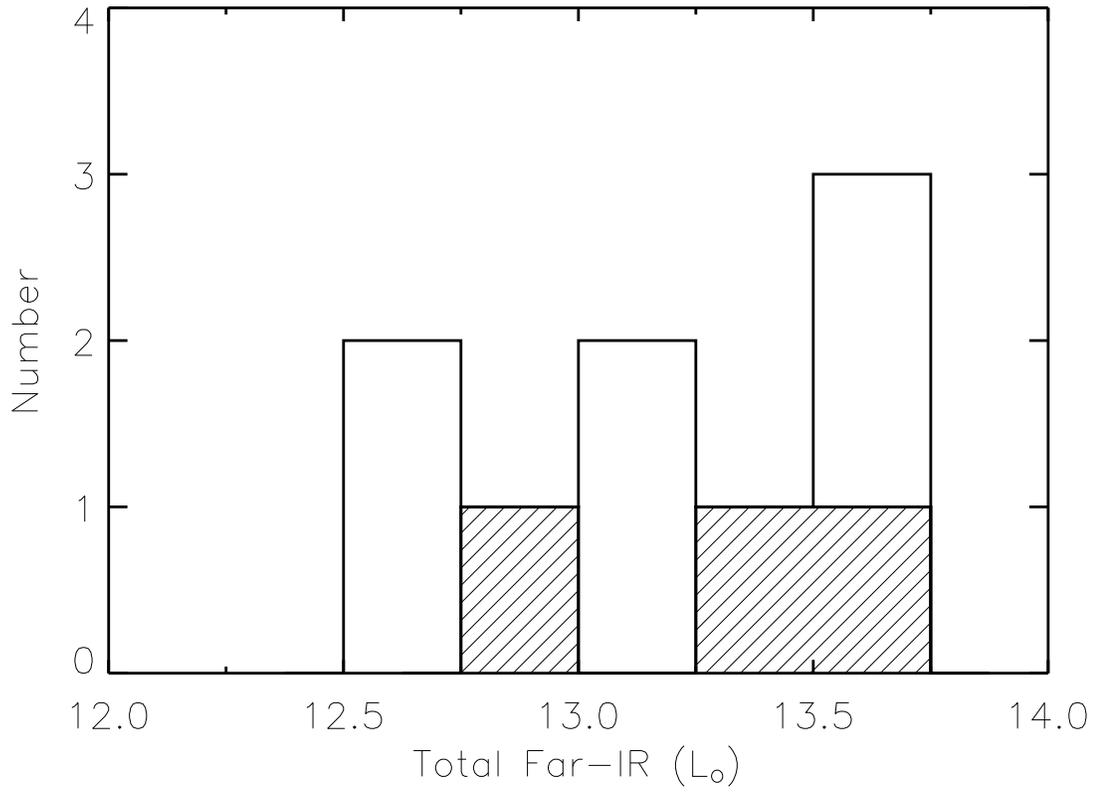}
\caption{Histogram of inferred total far-infrared luminosity for MIPS sources associated with $z$=2.38 
\Lya\ sources. The three sources potentially associated with the \Lya blobs B6 and B7 are marked with 
cross-hatching.}
\end{figure}

\begin{figure}
\plotone{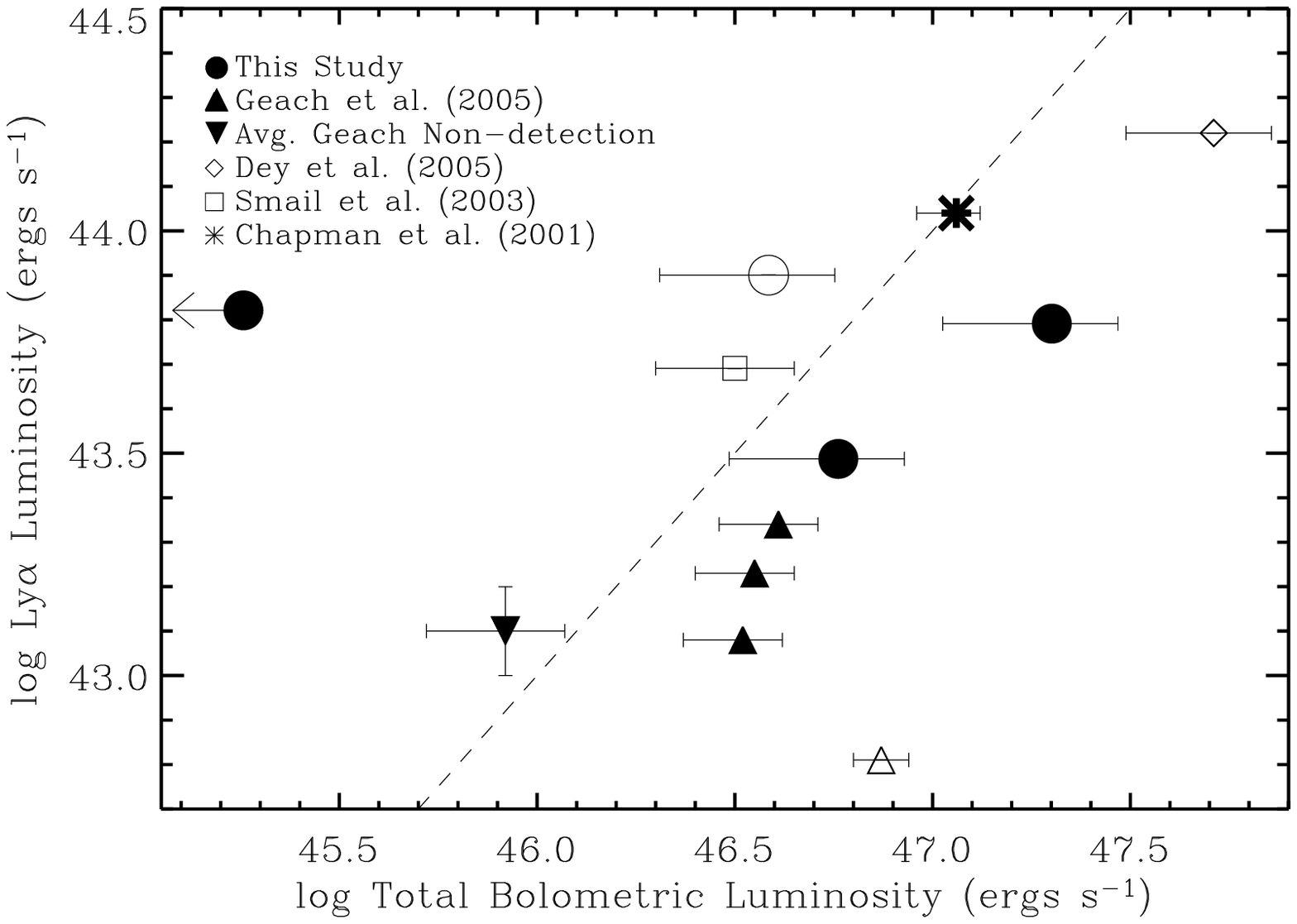}
\caption{Log L$_{Ly\alpha }$ vs. log L$_{bol}$ for high-z \Lya blob ULIRGs. The
24\m -detected blobs are the circles (This study) and the diamond \citep{dey05}, with
error bars based on the estimated reliability of the \cite{cha01} relation.  The 
submm detections are the triangles \citep{gea05}, squares \citep{sma03}, 
and asterisks \citep{chp01}. The upside-down triangle is the average of all non-detected Geach et al. blobs.  
\Lya blobs with evidence for AGN activity (X-rays or CIV emission) are plotted as hollow symbols.
The dashed line marks where L$_{Ly\alpha }$/L$_{bol}$ = 0.001.} 
\end{figure}

\end{document}